\documentstyle[prd,aps]{revtex}
\newcommand{\half}{{\textstyle{1\over2}}}
\newcommand{\tfrac}[2]{\textstyle{#1\over#2}}
\newcommand{\ma}[2]{{(\bar{#1}\,#2)}}
\newcommand{\mb}[2]{{{#1}\otimes{#2}}}
\newcommand{\mc}[3]{{{#1}\otimes{#2}\otimes{#3}}}
\newcommand{\md}[4]{{{#1}\otimes{#2}\otimes{#3}\otimes{#4}}}
\newcommand{\me}[3]{{(\bar{#1}\,#2\,#3)}}
\newcommand{\mee}[2]{{(\bar{#1}\,#2\,#1)}}
\newcommand{\id}{{\bf 1}}
\newcommand{\g}{\gamma}
\newcommand{\h}{\gamma_5}
\newcommand{\hg}{\gamma_5\gamma}
\newcommand{\s}{\sigma}
\newcommand{\hs}{\gamma_5\sigma}

\begin{document}
%\draft
\title{Calculating the Fierz Transformation for Higher Orders}
\author{J.~A.~Maruhn and  T.~B\"urvenich}
\address{Institut f\"ur Theoretische Physik, Universit\"at Frankfurt,
P. O. Box 111932, 60054 Frankfurt, Germany}
\author{D.~G.~Madland}
\address{Theoretical Division, Los Alamos National Laboratory,
Los Alamos, New Mexico 87545, USA}
\maketitle
\begin{abstract}
We consider the higher-order Fierz transformation, which
corresponds to expanding a product of $\bar\psi\Gamma\psi$ terms into
a sum of products of Dirac densities and currents. It is shown that
the Fierz transformation can be obtained by solving a large system of
linear equations with fractional complex coefficients, which is
practical at least up to fourth power.
\end{abstract}
\section{The problem}
\textit{Fierz transformation\/} \cite{fierz} is a name given to
the expression of
a certain product of nondiagonal matrix elements of Dirac
$\Gamma$-matrices as an expansion into products of diagonal matrix
elements, such as
\begin{equation}\label{base}
\me{a}{\Gamma_i}{b}\me{b}{\Gamma_j}{a}=\sum_{k,l=1}^{16}c_{kl}
\mee{a}{\Gamma_k}\mee{b}{\Gamma_l}.
\end{equation}
Here $\Gamma_i$ stands for one of the 16 Dirac matrices
$\{\id,\h,\g_\mu, \hg_\mu,\s_{\mu\nu}\}$ constituting a linearly
independent basis in the space of complex $4\times4$ matrices.
The matrix elements denote products in Dirac-index space only,
i.~e.,
\begin{equation}
\me{a}{\Gamma_i}{b}=\bar\psi_a({\bf r},t)\,
\Gamma_i\,\psi_b({\bf r},t).
\end{equation}
The Fierz transformation is useful for expressing exchange matrix
elements in terms of densities, currents, and other diagonal ones,
which greatly eases their use in, for example, relativistic
mean-field theories. For this reason it has been studied extensively,
see, e.~g., the papers by Y.~Takahashi \cite{takahashi} or
generalizations to SU(n) \cite{zong}.
 
In the context of nonlinear self-coupling of meson fields,
higher-order  versions of the Fierz transformation have
become of interest (for a review see \cite{hoch}).
If we express the order as the number of
$\Gamma$-matrices involved, the above Eq.\ (\ref{base}) is of second
order, and in this paper we will be concerned with constructing
expansions in third
\begin{equation}
\me{a}{\Gamma_i}{b}\me{b}{\Gamma_j}{c}\me{c}{\Gamma_k}{a}=\sum_{l,m,n=1}^{16}c_{lmn}
\mee{a}{\Gamma_l}\mee{b}{\Gamma_m}\mee{c}{\Gamma_n}
\end{equation}
and fourth order:
\begin{equation}
\me{a}{\Gamma_i}{b}\me{b}{\Gamma_j}{c}\me{c}{\Gamma_k}{d}\me{d}{\Gamma_l}{a}=
\sum_{m,n,p,q=1}^{16}c_{mnpq}
\mee{a}{\Gamma_m}\mee{b}{\Gamma_n}\mee{c}{\Gamma_p}\mee{d}{\Gamma_q}.
\end{equation}
Here, a general computer-based solution method will be discussed,
using either computer algebra systems such as \textit{Mathematica}
\cite{mathematica} or conventional programming allowing exact
manipulation of fractional complex numbers. Although this approach is
quite general, it will be applied preferentially to the case of
symmetrized matrix elements with only the matrix $\id$ appearing in
the expression to be expanded.
 
\section{Symmetrization}
In many applications, the wave function indices will all be summed
over, so that it is sufficient to deal with an expression symmetrized
over the indices. Thus, on the left-hand side of Eq.\ (\ref{base}) in
the second-order case we can write
\begin{equation}
\sum_{ab}\me{a}{\Gamma_i}{b}\me{b}{\Gamma_j}{a}=\half\sum_{ab}\left[
\me{a}{\Gamma_i}{b}\me{b}{\Gamma_j}{a}+\me{b}{\Gamma_i}{a}
\me{a}{\Gamma_j}{b}\right],
\end{equation}
and the right-hand side of Eq.\ (\ref{base}) will be symmetrized
exactly in the same way.
 
Using the notation $\sum_{\{abc\ldots\}}$ to refer to the sum over all
permutations of the symbols $a,b,c\ldots$, we can reformulate the
Fierz transformation
problem for the symmetrized matrix element as
\begin{equation}
\sum_{\{ab\}}\me{a}{\Gamma_i}{b}\me{b}{\Gamma_j}{a}=\sum_{\{ab\}}\sum_{kl}c_{kl}
\mee{a}{\Gamma_k}\mee{b}{\Gamma_l},
\end{equation}
where the factor $\half$ has been dropped on both sides.
 
It is important to realize that because of the product structure, the
right-hand side is symmetric under an exchange of the $\Gamma$
matrices as well. It is thus useful to introduce a notation for
symmetrized terms,
\begin{equation}
\mb{\Gamma_k}{\Gamma_l}=
\sum_{\{ab\}}\mee{a}{\Gamma_k}\mee{b}{\Gamma_l}
=\mee{a}{\Gamma_k}\mee{b}{\Gamma_l}+
\mee{a}{\Gamma_l}\mee{b}{\Gamma_k},
\end{equation}
which can easily be generalized to higher order, for example,
in third order the symmetrized problem becomes
\begin{equation}\label{symm1}
\sum_{\{abc\}}\me{a}{i}{b}\me{b}{j}{c}\me{c}{k}{a}=\sum_{l\leq
m\leq n}
c_{lmn}\mc{\Gamma_l}{\Gamma_m}{\Gamma_n}
\end{equation}
with
\begin{equation}\label{symm2}
\mc{\Gamma_l}{\Gamma_m}{\Gamma_n}=
\sum_{\{abc\}}\me{a}{\Gamma_l}{a}\me{b}{\Gamma_m}{b}\me{c}{\Gamma_n}{c},
\end{equation}
where symmetrization could equivalently be carried out in the
indices $l,m,n$ instead of $a,b,c$.
Terms of fourth and higher orders are defined analogously.
 
\section{Algebraic determination of the expansion}
The second-order Fierz transformation as defined in Eq.\ (\ref{base})
can be viewed as a system of equations obtained by comparing
coefficients in the $4^4=16^2$ dimensional space spanned by the spinors
$\psi_a$, $\psi_b$, $\bar\psi_a$, and $\bar\psi_b$. The coefficients
are given by the components of the $\Gamma$-matrices and thus can be
expressed as complex integers. The unknowns $c_{kl}$ are $16\times16$
in number, so that we have exactly the right number of equations, and
since the $\Gamma$-matrices form a basis for the $4\times4$ complex
matrices, the decomposition (\ref{base}) is always possible.
 
The solution of this system of linear equations can be carried out
using the standard Gauss elimination algorithm, provided that the
coefficients are not treated in floating point arithmetic, but as
exact complex fractions.
 
For third-order Fierz transformations the dimension of the system of
equations is $16^3$ and it is $16^4$ for fourth order, that is, the
complexity in going from second order to fourth order is in the ratio
1:16:256. For the latter case practical solution would require
substantial computing resources, but fortunately in cases of
practical interest the number of terms in the expansion can be
reduced substantially by symmetry and invariance requirements. In the
symmetrized case of the preceeding section, for example, the
dimension in fourth order is reduced by the number of permutations 4!.
 
\section{Selection of the terms in the expansion}
The expansion into products of the diagonal matrix elements of the
$\Gamma$-matrices is always possible, but usually is not the most useful
expression of the Fierz transformation. To see this, let us look at
an important special case: that of identity matrices on the left-hand
side. The decomposition problem thus is
\begin{equation}
\ma{a}{b}\ma{b}{a}=\sum_{jk}c_{jk}\me{a}{\Gamma_j}{a}\me{b}{\Gamma_k}{b},
\end{equation}
or, in symmetrized form,
\begin{equation}
\sum_{\{ab\}}\ma{a}{b}\ma{b}{a}=\sum_{jk}c_{jk}\mb{\Gamma_j}{\Gamma_k}.
\end{equation}
 
Since the left-hand side is a Dirac scalar, this means that the
right-hand side also can contain only scalar combinations of
$\Gamma$-matrices.  The only scalar combinations built out of
products of two $\Gamma$-matrices are $\mb\id\id$, $\mb\h\h$,
$\mb{\g_\mu}{\g^\mu}$, $\mb{\hg_\mu}{\hg^\mu}$, and
$\mb{\s_{\mu\nu}}{\s^{\mu\nu}}$, assuming the familiar index
summation convention. The Fierz transformation problem in this case
thus can be restated as (note that here because of the complete
symmetry of all terms, the symmetrization can be omitted):
\begin{eqnarray}\label{fierz4}
\ma{a}{b}\ma{b}{a}&=&c_1
\ma{a}{a}\ma{b}{b}+c_2\me{a}{\h}{a}\me{b}{\h}{b}
+c_3\me{a}{\g_\mu}{a}\me{b}{\g^\mu}{b}\\
&&+c_4\me{a}{\hg_\mu}{a}\me{b}{\hg^\mu}{b}
+c_5\me{a}{\s_{\mu\nu}}{a}\me{b}{\s^{\mu\nu}}{b}.
\nonumber
\end{eqnarray}
 
Note that symmetrization works slightly differently in this case: the
scalar products sometimes make certain index combinations appear
repeatedly in the expansion, but it is still sufficient to include
only one ordering of the $\Gamma$ matrices in the symmetrized terms.
 
Eq.\ (\ref{fierz4}) corresponds to 256 equations for the 5 unknown coefficients.
Clearly most equations will be redundant; eliminating them from the
statement of the problem, however, turns out to complicate the
solution, but the high degree of redundancy provides a welcome check
for completeness and consistency of the assumed decomposition.
 
\section{Solution for the fourth-order case}
Equation (\ref{fierz4}) will be used to illustrate the method of
solution. Actually two different approaches were used depending on
the complexity of the problem.
 
For second and third order a very simple but flexible approach was
implemented in \textit{Mathematica}. The spinors were expressed as
four-component vectors containing symbols of the form
\begin{equation}
\psi_a\rightarrow(a1,a2,a3,a4),\qquad
\bar\psi_a\rightarrow(aa1,aa2,aa3,aa4),\qquad
\psi_b\rightarrow(b1,b2,b3,b4),\qquad
\bar\psi_b\rightarrow(bb1,bb2,bb3,bb4),
\end{equation}
The scalar products with the $\Gamma$-matrices can then be evaluated
straightforwardly, yielding a representation of Eq.\ (\ref{fierz4}) as a
linear equation in the coefficients $c_i$ with coefficients
biquadratic in the spinor components.  The symmetrization is done
automatically. The coefficients are determined successively by
choosing one term in which $c_1$ occurs; if this is, e.~g.,
$c_1\,aa1\,a2\,bb2\,b4$, the coefficient of $aa1\,a2\,bb2\,b4$ is
extracted from the equation, yielding a linear equation in the $c_i$
alone, which is solved for $c_1$. This is inserted into the equation,
reducing the number of unknown coefficients by one, and the process
is repeated until all $c_i$ have been found. If any terms are then
left in the equation, the expansion was not complete.
 
While this is a very straightforward and not particularly efficient
solution, the steps can be automated using built-in functions, and it
is quite flexible, since new terms can be added by simply writing
them down as symbolic expressions in the spinors and
$\Gamma$-matrices. For the fourth order, however, this process turned
out to be too inefficient, so that as an alternative a Fortran-90
code was developed that uses a data type for fractional complex
numbers and straightforward Gauss elimination. Programming the
individual terms, though, requires substantially more coding.
 
Specifically, the four independent components of each spinor are
represented by a number $0,1,2,3$, corresponding to a two-bit
integer. The four spinor indices are then combined into an 8-bit
index, which indicates the index of the equation to which this
combination contributes. For example, the expression
$\ma{a}{b}\ma{b}{a}$ has (as only one of the nonvanishing matrix
elements) a value of one for the 3-component of
spinors $\bar\psi_a$ and $\psi_b$ and the 2-component of
spinors $\bar\psi_b$ and $\psi_a$. If we arrange the index with
$\bar\psi_a$, $\psi_a$, $\bar\psi_b$, $\psi_b$ in order of decreasing
magnitude, the index for the equation will be
\begin{equation}
3\times4^3+2\times4^2+2\times4+3=11101011_2=235.
\end{equation}
The program will loop through all spinor component combinations and
add the generated coefficients at the corresponding position in the
system of equations.
 
To help with the proper selection of terms in the expansion, the
code checks whether any of the coefficients of the term being
generated are nonzero, since symmetry may cause cancellation in a
non-obvious way, and also whether a term is directly proportional to
one previously generated.
 
The system of equations is then solved using Gauss elimination, where
more general linear dependences will become apparent. If the
elimination does not solve \textit{all\/} of the equations, it is clear
that the decomposition is not complete and this is then indicated.
 
\section{Results}
As sample results, we give the decomposition of the symmetrized term
in second, third, and fourth order for the case of identity matrices, as
tables of terms with the corresponding expansion coefficients. For
the higher-order terms, some remarks about new features are made.
 
\subsection{Second order}
The result for second order is
\begin{eqnarray}
\ma{a}{b}\ma{b}{a}&=&\tfrac1{4}
\ma{a}{a}\ma{b}{b}+\tfrac1{4}\me{a}{\h}{a}\me{b}{\h}{b}
+\tfrac1{4}\me{a}{\g_\mu}{a}\me{b}{\g^\mu}{b}\\
&&-\tfrac1{4}\me{a}{\hg_\mu}{a}\me{b}{\hg^\mu}{b}
+\tfrac1{8}\me{a}{\s_{\mu\nu}}{a}\me{b}{\s^{\mu\nu}}{b}.
\nonumber
\end{eqnarray}
This result is repeated in Table\ \ref{secondorder} in order to
indicate the correct reading of the tables for higher order, where,
however, a full symmetrization of both sides of the equation becomes
necessary according to Eqs.\ (\ref{symm1}) and (\ref{symm2}).
 
\subsection{Third order}
In third order the symmetrization is no longer trivial as it was in the
second-order case. It may be surprising that terms with $\hs_{\mu\nu}$
must be included; these can be equivalently formulated using the
identity
\begin{equation}
\hs_{\mu\nu}=
\tfrac{{\rm i}}{2}\epsilon_{\kappa\lambda\mu\nu}\sigma^{\kappa\lambda},
\end{equation}
but retaining the matrix $\h$ makes the space-reversal properties of the
terms more readily apparent. Note that either way our basis still consists
of only 16 linearly independent matrices.
The resulting transformation is given in Table\ \ref{thirdorder}.
 
\subsection{Fourth order}
In fourth order the number of possible terms becomes larger and it is
difficult to see which are independent. Using the basic building
blocks $\id$, $\h$, $\g_\mu$, $\hg_\mu$, $\s_{\mu\nu}$, and
$\hs_{\mu\nu}$ in all possible combinations fulfilling the condition
of coupling to a Dirac scalar, which also implies an even number of
$\h$-matrices, proves sufficient, but also leads to many dependent
terms which are eliminated with the program's help. The final
result is given in Table\ \ref{fourthorder}.
 
\section{Other applications}
 
A similar method can be applied to nonsymmetric and nonscalar terms;
in this case of course more terms will appear in the expansion. They
can be constructed as before from the basic building blocks $\id$,
$\h$, $\g_\mu$, $\hg_\mu$, $\s_{\mu\nu}$, and $\hs_{\mu\nu}$ in a way
that yields the desired Lorentz transformation properties.
 
As an example, the following decomposition was obtained:
\begin{eqnarray}\nonumber
\me{a}{\s_{\mu\nu}}{b}\me{b}{\g^\nu}{a}&=&\\ \nonumber
&&-\tfrac1{4}\mee{a}{\g^\nu}\mee{b}{\s_{\mu\nu}}
-\tfrac1{4}\mee{a}{\s_{\mu\nu}}\mee{b}{\g^\nu}\\
&&+\tfrac{3\rm{i}}4 \mee{a}{\g_\mu}\ma{b}{b}
-\tfrac{3\rm{i}}4 \ma{a}{a}\mee{b}{\g_\mu}\\ \nonumber
&&+\tfrac{3\rm{i}}4 \mee{a}{\hg_\mu}\mee{b}{\h}
+\tfrac{3\rm{i}}4 \mee{a}{\h}\mee{b}{\hg_\mu}\\ \nonumber
&&-\tfrac1{4} \mee{a}{\hg^\nu}\mee{b}{\hs_{\mu\nu}}
+\tfrac1{4} \mee{a}{\hs_{\mu\nu}}\mee{b}{\hg^\nu}.
\end{eqnarray}
Those terms where the exchange of indices $a$ and $b$ changes the sign
will drop out in the symmetrized version of this result, which is
\begin{eqnarray}
\half\left[\me{a}{\s_{\mu\nu}}{b}\me{b}{\g^\nu}{a}+
\me{b}{\s_{\mu\nu}}{a}\me{a}{\g^\nu}{b}\right]&=&
-\tfrac1{4}\mee{a}{\g^\nu}\mee{b}{\s_{\mu\nu}}
-\tfrac1{4}\mee{a}{\s_{\mu\nu}}\mee{b}{\g^\nu}\\
&&+\tfrac{3\rm{i}}4 \mee{a}{\hg_\mu}\mee{b}{\h}
+\tfrac{3\rm{i}}4 \mee{a}{\h}\mee{b}{\hg_\mu}. \nonumber
\end{eqnarray}
 
\section{Summary}
Once it is realized that constructing a Fierz transformation of any order
essentially just means solving a linear system of equations with
fractional complex coefficients, standard techniques of numerical
analysis are sufficient to solve the problem. The calculation is
now feasible up to fourth order, and the main work that
remains is the expression of the transformation through meaningful
couplings of the $\Gamma$ matrices, which still requires treating
each transformation separately. Our work here demonstrates
that tractable solutions may be possible for each of these cases.
 
Commented \textit{Mathematica} and Fortran-90 programs are available
from the authors.
 
\section{Acknowledgements}
This work was supported by the Bundesministerium f\"ur Bildung und
Forschung, by the Gesellschaft f\"ur Schwerionenforschung, and by
the U.S. Department of Energy under contract W-7405-ENG-36. The
authors are grateful to J. Reinhard and S. Schramm for critical
comments.

\narrowtext
\begin{table}
\caption{Terms and expansion coefficients appearing in second
order.} \label{secondorder}
\begin{tabular}{lr}
{\rm Term}&{\rm Coefficient}\\
\hline
$\mb{\id}{\id}$&$\tfrac1{4}$\\
$\mb{\h}{\h}$&$\tfrac1{4}$\\
$\mb{\g_\mu}{\g^\mu}$&$\tfrac1{4}$\\
$\mb{\hg_\mu}{\hg^\mu}$&$-\tfrac1{4}$\\
$\mb{\s_{\mu\nu}}{\s^{\mu\nu}}$&$\tfrac1{8}$\\
\end{tabular}
\end{table}
\begin{table}
\caption{Terms and expansion coefficients appearing in third
order.} \label{thirdorder}
\begin{tabular}{lr}
{\rm Term}&{\rm Coefficient}\\
\hline
$\mc{\id}{\id}{\id}$&$\tfrac1{16}$\\
$\mc{\id}{\h}{\h}$&$\tfrac3{16}$\\
$\mc{\id}{\g_\mu}{\g^\mu}$&$\tfrac3{16}$\\
$\mc{\id}{\hg_\mu}{\hg^\mu}$&$-\tfrac3{16}$\\
$\mc{\id}{\s_{\mu\nu}}{\s^{\mu\nu}}$&$\tfrac3{32}$\\
$\mc{\g_\mu}{\hg_\nu}{\hs^{\mu\nu}}$&$-\tfrac{3\textrm i}{8}$\\
$\mc{\h}{\s_{\mu\nu}}{\hs^{\mu\nu}}$&$\tfrac3{32}$\\
\end{tabular}
\end{table}
 
\begin{table}
\caption{Terms and expansion coefficients appearing in fourth
order} \label{fourthorder}
\begin{tabular}{lr}
{\rm Term}&{\rm Coefficient}\\
\hline
$\md{\id}{\id}{\id}{\id}$&$\tfrac1{64}$\\
$\md{\id}{\id}{\h}{\h}$&$\tfrac3{32}$\\
$\md{\id}{\id}{\g_\mu}{\g^\mu}$&$\tfrac3{32}$\\
$\md{\id}{\id}{\hg_\mu}{\hg^\mu}$&$-\tfrac3{32}$\\
$\md{\id}{\id}{\s_{\mu\nu}}{\s^{\mu\nu}}$&$\tfrac3{64}$\\
$\md{\id}{\h}{\s_{\mu\nu}}{\hs^{\mu\nu}}$&$\tfrac3{32}$\\
$\md{\id}{\g_\mu}{\hg_\nu}{\hs^{\mu\nu}}$&$-\tfrac{3\textrm i}{8}$\\
$\md{\h}{\h}{\h}{\h}$&$\tfrac1{64}$\\
$\md{\h}{\h}{\g_\mu}{\g^\mu}$&$\tfrac1{32}$\\
$\md{\h}{\h}{\hg_\mu}{\hg^\mu}$&$-\tfrac1{32}$\\
$\md{\h}{\h}{\s_{\mu\nu}}{\s^{\mu\nu}}$&$\tfrac3{64}$\\
$\md{\h}{\g_\mu}{\hg_\nu}{\s^{\mu\nu}}$&$-\tfrac{1\textrm i}{8}$\\
$\md{\g_\mu}{\g^\mu}{\g_\nu}{\g^\nu}$&$\tfrac1{64}$\\
$\md{\g_\mu}{\g^\mu}{\hg_\nu}{\hg^\nu}$&$-\tfrac3{32}$\\
$\md{\g_\mu}{\g^\nu}{\hg_\mu}{\hg^\nu}$&$\tfrac1{16}$\\
$\md{\g_\lambda}{\g^\lambda}{\s_{\mu\nu}}{\s^{\mu\nu}}$&$\tfrac3{64}$\\
$\md{\g_\lambda}{\g^\mu}{\s_{\mu\nu}}{\s^{\lambda\nu}}$&$-\tfrac1{16}$\\
$\md{\hg_\mu}{\hg^\mu}{\hg_\nu}{\hg^\nu}$&$\tfrac1{64}$\\
$\md{\hg_\lambda}{\hg^\lambda}{\s_{\mu\nu}}{\s^{\mu\nu}}$&$-\tfrac3{64}$\\
$\md{\hg_\lambda}{\hg^\mu}{\s_{\mu\nu}}{\s^{\lambda\nu}}$&$\tfrac1{16}$\\
$\md{\s_{\mu\nu}}{\s^{\mu\nu}}{\s_{\kappa\lambda}}{\s^{\kappa\lambda}}$&$\tfrac3{256}$\\
$\md{\s_{\mu\nu}}{\s^{\lambda\nu}}{\s_{\lambda\kappa}}{\s^{\mu\kappa}}$&$-\tfrac1{64}$\\
\end{tabular}
\end{table}
\end{document}